\documentclass[twocolumn,prl]{revtex4}
\usepackage{graphicx}
\usepackage{amsmath}

\begin{document}
\title{Influence of a Mesoscopic Bath on Quantum Coherence}
\author{Onuttom Narayan$^1$}
\author{Harsh Mathur$^2$}
\affiliation{$^1$ Department of Physics, University of California, 
Santa Cruz, CA 95064\\
$^2$ Physics Department, Case Western Reserve University, 
10900 Euclid Avenue, Cleveland OH 44106-7079}

\begin{abstract}
For a quantum double well system interacting with a mesoscopic
bath, it is shown that a single particle in the bath is sufficient to
substantially reduce tunneling between the two wells. This is demonstrated
by considering an ammonia molecule in the center of a ring; in addition
to halving the maser line frequency, there is an increase in intensity
by four orders of magnitude. The tunneling varies non-monotonically with the
number $N$ of electrons in the ring, reflecting the changing electronic
correlations. Although the tunneling is reduced for
small $N$, it turns around and grows to its free value for large
$N.$ This is shown to not violate Anderson's orthogonality theorem.
Experimental implementations are discussed.

\end{abstract}

\maketitle
The importance of decoherence for the transition from quantum to
classical behavior is well known. Various authors have shown that
when a system is coupled to a bath consisting of an infinite number
of degrees of freedom, irreversibility---and therefore, 
decoherence---is a consequence; this has given rise to the field of dissipative
quantum mechanics\cite{caldeira,leggett}.  Studies in this field
have been further stimulated by the experimental realisation
of quantum coherent mesoscopic systems that are small on the macroscopic
scale but large on the microscopic, atomic scale\cite{meso}.  Control of
coupling to the environment is also critical to the success of quantum
computing, a fast-expanding area of research\cite{chuang}.

In this paper we consider a microscopic quantum system that is
predominantly coupled to a {\it mesoscopic\/} bath. In dissipative
quantum mechanics, it is generally presumed that the quantum system
is coupled to a macroscopic bath with an essentially infinite number
of degrees of freedom. The general belief is that ``more is worse'':
the greater the coupling to the environment, the more quantum effects
are destroyed. Here we show that, for a mesoscopic bath, it is possible
for the tunneling between the two wells of a quantum system in a double
well potential to be highly non-monotonic as a function of the number of
particles in the bath. There is a substantial reduction in the tunneling
with a {\it single\/} particle. As the number of particles is increased,
the effect of the mesoscopic bath at first sharply decreases, then grows,
and finally decreases again, essentially vanishing for sufficiently
large particle number.  We will show that the last assertion does not 
contradict Anderson's orthogonality catastrophe~\cite{anderson}.

We demonstrate these results by considering an often studied double well
system, the ammonia molecule\cite{feynman}, at the center of
a mesoscopic ring. The mesoscopic bath is taken to be a one
dimensional ring; qualitatively similar results
would be obtained for a multichannel ring or for a singly connected
geometry such as a disk.  If the ring has just one electron, the tunneling
between the two configurations for the ammonia dipole is reduced by as
much as 50\%, reducing the frequency of the transition
between the symmetric and antisymmetric configurations by the same
amount. In addition, the coupling to the electron in the ring increases
the dipole moment approximately hundredfold. This dramatically enhances
the intensity of the spectral line for the transition, by 
$\sim 10^4.$ As the number of electrons on the ring is increased, the
rich non-monotonic behaviour noted above unfolds.  Possible experimental
realizations are discussed towards the end of the paper.

Parenthetically we note that it is known the coupling of a
molecule to light is enhanced by proximity to a metallic nanoparticle,
an effect that is the basis of sensitive optical detection of molecules
\cite{vanduyne}. The physical origin of this essentially classical
effect is field enhancement due to the coupling of optical radiation
to metallic plasmons. In contrast the enhancement in the spectral
line identified here is quantum in origin.

To establish these results, first consider an ammonia molecule in the
center of a ring of radius $R.$ Assume for the moment that the dipole
moment of the molecule is in the plane of the ring, and that there is
only one electron in the ring.  In the absence of the ring, for every
orientation of the hydrogen plane, there are two possible positions of
the nitrogen atom. The ground and first excited states are the symmetric
and antisymmetric combinations of these. Transitions between the two
states are used in the ammonia maser.

In the presence of the ring, the two possible positions of the nitrogen
atom, which have opposite dipole moments, perturb the electron in the
ring, changing its ground state to two different ground states. The
possible states of the combined system of the ammonia molecule and the
electron in the ring are then
\begin{equation}
|\uparrow\rangle\otimes|E_n\rangle,\qquad 
|\downarrow\rangle\otimes|E_n^\prime\rangle
\end{equation}
where $|E_n\rangle$ is the $n^{{\rm th}}$ electronic eigenstate when the ammonia
electric dipole moment is in one orientation, and $|E_n^\prime\rangle$ is the
corresponding electronic eigenstate when the dipole moment is in the
other orientation. The fact that $|E_n\rangle$ and $|E_n^\prime\rangle$
are different---partially orthogonal---states reduces the tunneling
between the two dipole orientations, since the tunneling part of
the Hamiltonian connects $|\uparrow\rangle\otimes|\psi\rangle$ to
$|\downarrow\rangle\otimes|\psi\rangle$ with the same $|\psi\rangle$ for
both states. Thus the splitting between the symmetric and antisymmetric
state is reduced. The different orientations of the hydrogen triangle
do not affect this calculation, since there is no tunneling between the
different orientations, only between the two different positions of the
nitrogen atom for a fixed orientation of the hydrogen triangle.

The potential of the electron in the ring due to the ammonia dipole moment
is $V(\theta) = e p \cos \theta / (4 \pi \epsilon_0 R^2)$,
causing matrix elements between the unperturbed
electronic eigenstates. The unperturbed eigenstates of the electron
are $\psi_0 = 1/\sqrt{2\pi}$ and 
$\psi_n(\theta) = \cos(n \theta)/\sqrt\pi$ for $n\neq 0,$ 
with energies 
$\hbar^2 n^2/(2 m R^2).$ We neglect the $\sin(n\theta)$
wavefunctions, since they are not mixed into the ground state by
$V(\theta).$ The Hamiltonian for the electron in the presence
of the ammonia molecule is 
\begin{equation}
H_{ij} = {{\hbar^2}\over{2 m R^2}}
[j^2\delta_{ij} + \rho (\delta_{i, j+1} + \delta_{i, j-1}) 
+ (\sqrt 2 - 1) \rho (\delta_{i,1}\delta_{j, 2} + \delta_{i,2}\delta_{j,1})]
\label{elham}
\end{equation}
where $\rho =  m e p/(4 \pi \epsilon_0 \hbar^2) $
is the dimensionless perturbation strength. Using
$\rho\approx 0.6137$ D for ammonia, $H$
can be truncated to a $n\times n$ matrix, and its eigenvalues obtained
numerically. Changing $n$ from 4 to 5 changes the first
four eigenvalues by less than 1\%, so we use a $4\times 4$ truncation.
The eigenvalues are 
\begin{equation}
[E_0, E_1, E_2, E_3] = {{\hbar^2}\over{2 m R^2}}[-0.5233, 1.393, 4.055, 9.075].
\label{evalues}
\end{equation}
Including the configuration of the ammonia dipole, the Hamiltonian can
be written in the basis
$|\uparrow\rangle\otimes|E_n\rangle, |\downarrow\rangle
\otimes|E_n^\prime\rangle$
which is an $8\times 8$ matrix with our truncation. In the absence of
tunneling between the dipole configurations, this would be a diagonal
matrix, with doubly degenerate energy eigenvalues.

Without the electron in the ring, the splitting between the symmetric
and antisymmetric levels of the ammonia molecule
is $10^{-4}$ eV. In terms of the units used
in Eq.(\ref{elham}), for a ring of radius 10 nm, the tunneling matrix
element is thus 
\begin{equation}
t = -0.131 {{\hbar^2}\over{2 m R^2}}.
\label{tunneling}
\end{equation}
Since the tunneling Hamiltonian does not act on the electron
in the ring,  the tunneling matrix element between
$|\uparrow\rangle\otimes|E_n\rangle$ and 
$|\downarrow\rangle\otimes|E_m^\prime\rangle$ is 
$ t \langle E_m^\prime| E_n\rangle,$ 
which can be evaluated numerically. 

If $t$ is sufficiently small, the degenerate ground state in the $8\times
8$ Hamiltonian is split by $2t \langle E_0|E_0^\prime\rangle.$ 
Numerically, with Eq.(\ref{tunneling}),
{\it i.e.\/} a ring of radius 10 nm, the splitting between 
the lowest 
two eigenvalues is $0.122 \hbar^2/(2m),$ to a good approximation equal
to $2|t| \langle E_0|E_0^\prime\rangle.$
The frequency of the radiation for transition between these two
states is thus less than for the free ammonia molecule by a factor
of approximately 0.5. 
If the temperature is large compared to $(E_1 - E_0),$ one has to consider
transitions between excited states as well. For a 10 nm ring, the 
temperature $T$ has to be much less than $8 K$ to ignore the excited states. 
For $R < 30$ nm and $T << 8$ K, to a good approximation the Hamiltonian 
can thus be truncated to a symmetric $2\times 2$ matrix, for which the 
ground and first excited states are the symmetric and antisymmetric
combinations 
$|I\rangle, |II\rangle  = {1\over{\sqrt 2}}[|\uparrow; E_0\rangle 
\pm |\downarrow; E_0^\prime\rangle].$ 
In order to excite transitions between $|I\rangle$ and $|II\rangle,$
one has to apply a time dependent electric field, which couples to
the dipole moment of the states. With $P = P_{amm} + P_{el},$ one can
verify that 
$\langle I|P|II\rangle
= \langle \uparrow|P_{amm}|\uparrow\rangle + \langle E_0| P_{el}| E_0
\rangle.$ Since the electronic ground state has a dipole moment opposed
to the direction of the ammonia moment, this changes the intensity of
the spectral line. Numerically, one obtains 
\begin{equation}
\langle I|P|II \rangle = 0.325 e {\rm \AA} - 0.65 e R.
\label{dipole}
\end{equation}
For $R=10$ nm, $\langle I | P |II\rangle$ is changed 
by a factor of -200, increasing
the line intensity by $4\times 10^4.$ (Recall that $R < 30$ nm
for the $2\times 2$ truncation to be valid.)

We now reexamine the approximations made so far.  For a general
orientation of the ammonia molecule, only the component of its dipole
moment in the plane of the ring couples to the electron in the ring,
reducing the effects discussed above. 
For randomly oriented molecules, a broad
band will be seen.  Also, if the molecule is not in the center of the
ring, the symmetry between the two orientations of its dipole moment is
destroyed: the dipole moment prefers to align itself towards the closest
point on the ring. 

We now consider the effect of increasing the number of electrons in the 
ring. First, we consider the two electron case in detail. 
If $\theta_1$ and $-\theta_2$ are the angular positions of the electrons
with respect to the dipole moment of the ammonia, changing 
to $\theta_+ = (\theta_1 + \theta_2)/2$ and $\theta_- = (\theta_1 -
\theta_2),$ the electronic Hamiltonian is
\begin{equation}
H = {{\hbar^2}\over{4mR^2}}\Big[\partial_+^2 + 4 \partial_-^2\Big]
- 2 {{e p}\over{R^2}} \cos\theta_+\cos({{\theta_-}\over 2}) + {{e^2}\over
{2 R \cos\theta_-}}.
\label{hamilt}
\end{equation}
In the absence of the ammonia dipole, the Hamiltonian is
separable. Since the kinetic energy is small compared to the
potential energy, $\theta_-\approx \pi.$ Expanding $\theta_-$ as
$\pi - 2 \delta,$ $H_\delta \approx
(\hbar^2/mR^2)\partial^2/\partial_\delta^2 + e^2\delta^2/4 R,$ so that
\begin{equation}
\langle\delta^2\rangle = \hbar (m R e^2)^{-1/2}.
\label{deltarms}
\end{equation}
The wavefunction is uniform in the $\theta_+$ coordinate. 

With the dipole, the Hamiltonian of Eq.(\ref{hamilt}) is no longer
separable, so we perform an approximate analysis. From classical
arguments it is clear that $\langle\delta(\theta_+)\rangle$ is greatest
when $\theta_+=0,$ for which case by minimizing $-2 (e p/R^2)\sin\delta
+ e^2/(2 R \cos\delta)$ we find that $\langle\delta^2\rangle\approx
(4 p/e R)^2.$ By comparing with Eq.(\ref{deltarms}), we see that the
change in the ground state wavefunction is negligible for the $\delta$
coordinate. For the $\theta_+$ coordinate, with $\delta$ set at its
optimal value ($\delta = 4 p/e R$ for $\theta_+ = 0,$ $\delta = 0$
for $\theta_+ = \pm \pi/2$), there is an effective potential energy $
- (4 p^2/R^3) \cos^2\theta_+.$ Since this is negligible compared to
$\hbar^2/2mR^2,$ the kinetic energy still forces the wavefunction to
be essentially uniform in $\theta_+.$ Thus the overlap in the ground
state wavefunction of the two electrons with and without the ammonia
dipole potential is almost unity, as is therefore the overlap $\langle
E_0^\prime| E_0\rangle.$

If more electrons are added to the ring, $\langle E_0^\prime| E_0\rangle$
decreases: for $N$ electrons at low density the electronic state
is essentially a Wigner crystal and the dipole potential shifts each
electron from its lattice position by an angle $O(p/eRN),$ whereas
the angular spread of each electron wavefunction is $O(1/N).$ Thus
$\langle E_0\prime|E_0\rangle = [1 - O(p/eR)^2]^N,$ which tends to zero
for large $N.$ But in reality, as $N$ increases, the Wigner crystal
melts\cite{ceperley}. The overlap $\langle
E_0^\prime| E_0\rangle$ then {\it reverses\/} its $N$ dependence,
approaching unity as $N \rightarrow \infty$. Physically this is
because in the liquid state the electrons efficiently screen the dipole
potential so that it does not significantly perturb the electronic
ground state.

To make this plausible, let us consider the more tractable problem of 
$N$ electrons in a spherical shell of inner radius $R,$ perturbed by 
a point charge $q$ placed at the center of the shell. Within the Thomas
Fermi approximation, the point charge potential is screened within 
a distance $\sim d_s$ from $R.$ If $d_s<<R,$ screening also 
changes $V(R)$ from $q/R$ to  $q d_s/R^2.$ The phase shift 
$\delta_l(k_F)$ due to this potential is approximately zero for 
$l \gg k_F R,$ because of the centrifugal barrier, whereas for 
$l \ll k_F R,$ it is approximately $\delta_0(k_F).$  Within the WKB
approximation, $(\delta_0(k_F) \sim V(R)d_s/\hbar v_F).$
Following Anderson~\cite{anderson}, the overlap between the ground 
state wavefunction with and without the point charge is bounded by
$\exp[-\sum_l (2l + 1)\sin^2\delta_l(k_F)\ln N].$ Combining our results 
so far, the prefactor to $\ln N$ is 
$\sim -(k_F R)^2 (q d_s/R^2)^2 (d_s/\hbar v_F)^2.$ If $N$ is 
increased at fixed $R,$ since $1/d_s\sim k_F\sim N^{1/3},$ the 
overlap bound is $\exp[-\sim \ln N/N^{4/3}],$ which approaches unity as
$N\rightarrow\infty.$ This does not conflict with Anderson's 
result, which applies when $N$ and the system size are both increased,
with the particle density held constant.

Alternatively, we could consider $N$ electrons in a spherical box
of radius $R$ perturbed by a point charge $-e$ placed at the centre
of the box. Anderson's bound on the overlap can also be expressed 
as 
$ \exp \left( - k^2 \sigma_{{\rm tot}} \ln N/(12 \pi^3) \right)$
where $ \sigma_{{\rm tot}} $ is the cross section for an electron at
the Fermi surface to scatter from the screened potential of the
perturbing point charge. If we take the screened potential to be
of the Thomas-Fermi form and use the Born approximation,
both justified in the high density limit $k_f a_B \gg 1$
(where $a_B = 4 \pi \epsilon_0 \hbar^2 / m e^2 $ is the Bohr
radius), we find
\begin{equation}
\langle E_0^\prime| E_0 \rangle \leq \exp
\left[ - \frac{1}{12 \pi} \left( \frac{4}{9 \pi} \right)^{1/3}
\frac{1}{N^{1/3}} \frac{R}{a_B} \ln N \right]
\label{eq:bound}
\end{equation}
which tends to $1$ as $ N \rightarrow \infty$
if $R$ is held constant. Again, this does not conflict with
Anderson's result\cite{anderson}.

The case of a strictly one-dimensional ring deserves special
consideration. 
For spinless electrons with short-range interactions at high density,
the problem can be analysed by bosonisation, whereby
the electron liquid is described in terms of 
non-interacting bosonic density-wave oscillators\cite{vandelft}.
In this description, the smooth ammonia dipole
potential perturbs just one bosonic oscillator via a linear
coupling. Thus, in contrast to the case of a sharp Kane-Fisher
impurity\cite{fisher}, the problem remains trivially soluble.
A simple calculation reveals that the overlap of the electronic
ground states corresponding to the two configurations of the 
ammonia molecule is $\exp[-4 \rho^2 g/v^2 N^2],$ essentially
perfect as $N\rightarrow\infty.$ Here $v$ and $g$ are 
Luttinger liquid interaction parameters~\cite{vandelft}.
Physically, the reason for this dependence is that the single
oscillator to which the ammonia molecule couples becomes stiffer
as $N$ (and therefore the Fermi velocity) increases. 

In three dimensions, the single particle level spacing scales as
$1/N^{1/3}$. Thus as $N$ continues to grow,
the bath will ultimately cross over to a macroscopic regime when the
single particle level spacing falls well below the tunneling scale
$t$. In this regime, the ammonia molecule can couple to an enormous
number of possible particle-hole excitations of the bath, and its
tunneling behaviour should therefore be well-described in the conventional
framework of the spin-boson model\cite{leggett,remark}. In contrast to the
mesoscopic bath, the imperfect overlap between the ground states of a
macroscopic bath can renormalize the 
tunneling frequency of the two level system all
the way down to zero, leading to a complete suppression of tunneling, or
localization~\cite{leggett}.  Also, there is damping for the macroscopic
bath, but not the mesoscopic.  Another distinction is the mesoscopic
bath-induced amplification of the dipole coupling of ammonia to external
radiation. This has, to our knowledge, no counterpart in the physical
realisations of the spin-boson model studied so far.

We now briefly consider possible experimental realisations.
Metallic rings are unsuitable because of the high density of electrons
in metals: rings of the requisite size would contain far too many
electrons. Conventional semiconductor devices such as silicon MOSFETs
and GaAs MODFETs can have the required low electron density but suffer
from the difficulty that the electron gas is buried deep inside the
device below a dielectric layer of oxide or semiconductor. However it
may be possible to circumvent this difficulty by fabricating devices in
which the electron gas lies on the outer surface of the semiconductor
and in which the vacuum plays the role of the dielectric layer. GaAs is
unsuitable for such devices because of a high density of surface traps,
but Kane and co-workers have recently succeeded in passivating the
$\langle 111 \rangle$ surface of silicon and making working devices of
this type\cite{kane}. 
By suitably gating these devices, in principle, it should be
possible to laterally confine the electrons to rings. Other systems in
which the electron gas has low density and is not deeply buried include
electrons deposited on liquid helium\cite{cole} 
and newer semiconductor structures
that are grown by bottom-up techniques and that may be injected with
electrons by photo-excitation of a dye with which the structures have been
coated\cite{bawendi}. 
In the former system too electrons could be shepherded into rings
by suitable gating\cite{arnie}; 
in the latter, the electrons would be confined to
nanoscale semiconductor particles that are typically spherical in shape.

Apart from fabricating the rings, the other experimental task is to
position ammonia molecules in their vicinity. A simple extension of
the calculation above shows that a free ammonia molecule placed at the
center of a ring experiences a force that tends to deflect it off-center
where the coupling to the ring is much weaker. However this should
not be an insurmountable problem at least in the solid state systems
mentioned above. For example, with silicon devices, the
ammonia molecules could be deposited at random on the silicon surface, 
to which they would stick. Although ammonia sticks to a bare silicon
surface by covalent bonding of the nitrogen atom~\cite{stick} that renders
dipolar oscillations impossible, on a
hydrogen passivated Si surface such as used in the devices of Ref.\cite{kane}
it is expected that ammonia would stick without such covalent bonding.

In summary, we have shown that a quantum two level system coupled to
a mesoscopic bath responds to the bath in an extremely non-monotonic
manner as the number of particles in the bath is changed, and has a
strong response even with one particle in the bath. Apart from its 
intrinsic interest, such an interaction could be relevant to a quantum
computing architecture integrating mesoscopic solid state and flying atomic
or ionic qubits\cite{demarco}, a potential application deserving further 
exploration. As a prototype,
we have considered an ammonia molecule in the center of a ring with
a small number of electrons. We have discussed possible experimental
techniques to make such devices with a single electron in a ring.
The experimental signatures of the effects studied here are: (i) a
shift in the ammonia spectral line due to the suppressed tunneling,
(ii) a factor of $10^4$ increase in the strength of the line due to
the enhanced effective dipole moment and (iii) strong inhomogeneous
broadening of the line due to variations in the rings and the couplings
between the rings and ammonia molecules.
A key role in our analysis is played by the sensitivity of the 
electronic ground state to external perturbation. It is desirable to
determine the dependence of this ``wave function stiffness" on $N$
and $N/R$ more rigorously, e.g. numerically.

It is a pleasure to acknowledge helpful discussions with Mike Crommie,
Arnie Dahm, Josh Deutsch, Carine Edder, Anupam Garg, Bruce Kane,
Peter Littlewood and Jie Shan.

\end{document}